\documentclass[twocolumn,nofootinbib,floatfix,10pt,aps]{revtex4-2}
\usepackage{amsmath}
\usepackage{amssymb}
\usepackage{wasysym}
\usepackage{graphicx}
\usepackage{color,soul}
\usepackage{physics}
\usepackage{siunitx}
\usepackage{dsfont}
\usepackage{float}
\usepackage{multirow}
\usepackage[english]{babel}
\usepackage{blindtext}
\usepackage[english,nomargin,inline,marginclue,draft]{fixme}
\pdfpageheight\paperheight
\pdfpagewidth\paperwidth
\usepackage[colorlinks,linkcolor=blue,anchorcolor=blue,citecolor=blue,urlcolor=blue]{hyperref}
            
\fxusetheme{colorsig}
\FXRegisterAuthor{cg}{acg}{CG}  
\FXRegisterAuthor{th}{ath}{\color{blue}TH}  
\FXRegisterAuthor{ib}{aib}{\color{red}IB} 
\FXRegisterAuthor{sh}{ash}{\color{cyan}SH} 
\FXRegisterAuthor{db}{adb}{\color{green}DB} 
\FXRegisterAuthor{ps}{aps}{PS}
\makeatletter
\renewcommand*\FXLayoutInline[3]{%
  {\@fxuseface{inline}\ignorespaces{\color{fx#1}[#3: #2]}}}
\makeatother

\long\def\symbolfootnote[#1]#2{\begingroup%
\def\thefootnote{\fnsymbol{footnote}}\footnotetext[#1]{#2}\endgroup}

\def\nobreakbefore{%
  \relax\ifvmode\else
    \ifhmode
      \ifdim\lastskip > 0pt\relax
        \unskip\nobreakspace
      \else 
        \nobreakspace
      \fi
    \fi
  \fi
}
\let\oldcite\cite
\renewcommand\cite{\nobreakbefore\oldcite}

\begin{document}
\title{Rydberg microwave frequency comb spectrometer}

\author{Li-Hua Zhang}
\affiliation{Key Laboratory of Quantum Information, University of Science and Technology of China, Hefei, Anhui 230026, China.}
\affiliation{Synergetic Innovation Center of Quantum Information and Quantum Physics, University of Science and Technology of China, Hefei, Anhui 230026, China.}
\author{Zong-Kai Liu}
\affiliation{Key Laboratory of Quantum Information, University of Science and Technology of China, Hefei, Anhui 230026, China.}
\affiliation{Synergetic Innovation Center of Quantum Information and Quantum Physics, University of Science and Technology of China, Hefei, Anhui 230026, China.}
\author{Bang Liu}
\author{Zheng-Yuan Zhang}
\author{Guang-Can Guo}
\author{Dong-Sheng Ding}
\email{dds@ustc.edu.cn}
\author{Bao-Sen Shi}
\email{drshi@ustc.edu.cn}
\affiliation{Key Laboratory of Quantum Information, University of Science and Technology of China, Hefei, Anhui 230026, China.}
\affiliation{Synergetic Innovation Center of Quantum Information and Quantum Physics, University of Science and Technology of China, Hefei, Anhui 230026, China.}

\date{\today}

\begin{abstract}
Developing frequency combs spectral technologies has potential applications and prospects in the wide fields of cosmology, meteorology, and microwave measurement. Here, we demonstrate a Rydberg microwave frequency comb spectrometer via multiple-microwave field dressing, providing precise microwave measurement. The Rydberg microwave frequency comb spectrum provides a real-time and absolute frequency measurement with a range of 125 MHz and gives the relative phase of a single-frequency microwave signal. The reported experiment helps real-time sensing of an unknown microwave signal in a wide range using Rydberg atoms, which are vital for Rydberg-atom-based microwave metrology.
\end{abstract}
\maketitle
\section{Introduction}
A frequency comb consists of phase-stabilized narrow frequency lines with equidistant mode spacing (repetition frequency) \cite{hansch2006nobel}. Frequency comb spectroscopy has found various applications in frequency, time, and distance measurement over the past decade and in precision spectroscopy \cite{picque2019frequency} such as optical frequency synthesis and frequency reference in optical atomic clocks \cite{ma2004optical,RevModPhys.87.637}. There is much experimental progress on excitation of frequency comb spectroscopy in a variety of systems based on different physical mechanisms, such as a second-order Kerr optical frequency comb in a microresonator \cite{rueda2019resonant,lucas2020ultralow}, a broadband microwave (MW) frequency comb via the nonlinear pumped cavity in a superconductor resonator \cite{erickson2014frequency}, and a MW frequency comb generated in a circuit QED device \cite{PhysRevApplied.15.044031}. Other MW frequency comb generation methods include semiconductor lasers with harmonic frequency locking \cite{Chan,Juan} and optomechanical effects\cite{PhysRevLett.127.134301}.
Frequency comb spectroscopy with various systems and different bands such as the optical frequency band and the MW frequency band have distinct applications. For example, a MW band frequency comb spectrum could provide a method of directly measuring MW electric fields \cite{picque2019frequency}, detecting free radicals \cite{dousmanis1955microwave}, and monitoring the intermediate products of a chemical reaction \cite{hirota2012high}.

Exploring state-of-the-art microwave frequency comb (MFC) spectroscopy is important for such fields as timing, metrology, communications, and radio-astronomy. MFC spectroscopy with its wide range, low noise, and high stability plays a crucial part in MW applications \cite{picque2019frequency}. Compared with an optical frequency comb, an MFC can have small mode spaces across its bandwidth owing to the lower carrier frequency. Because of the large transition dipole moment \cite{saffman2010quantum}, the coupling between Rydberg atoms and the external MW electric fields is strong \cite{mohapatra2008giant,fan2015atom}. Developing a new type of Rydberg MFC spectrometer would have distinct significance because Rydberg atoms as a metrological resource have potential advantages in amplitude \cite{sedlacek2012microwave,facon2016sensitive,jing2020atomic}, phase \cite{simons2019rydberg,2021Aoa} sensing, channel capacity \cite{PhysRevLett.121.110502}, subwavelength imaging \cite{Fan:14}, have a wide working bandwidth \cite{Waveguide2021},  and the Rydberg atomic sensor can be combined with the deep learning technology \cite{liu2022deep}.

In this work, we theoretically and experimentally demonstrate a microwave sensing protocol via a Rydberg MFC method in a thermal cesium (Cs) atomic gas. Via multiple local MW fields, the excited Rydberg atoms are dressed and display comb-like RF transitions. We show the proof-of-principle effectiveness of this spectral detection of the MW fields. A signal MW field with a maximum frequency range of $\pm$ 3.5\,MHz is measured with a resolution of $\sim$ 0.1\,kHz, the amplitude and phase of the signal field are both retrieved and a measurement range of 125\,MHz is demonstrated. The Rydberg MFC method does work for a microwave signal with a certain spectral width. The demonstrated Rydberg MFC spectrometer is a first step towards precise microwave measurement using the atom-based MFC method.   

\section{method and experiment setup}

\begin{figure*}[t]
\includegraphics[width=2\columnwidth]{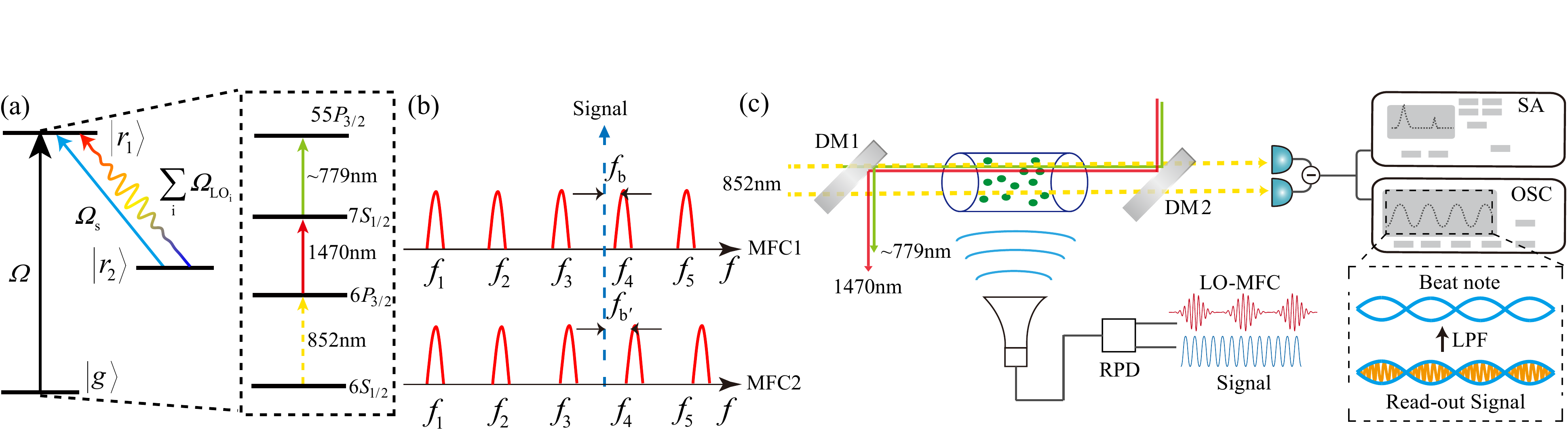}\caption{(a) The energy diagram of a Rydberg MFC spectrum. A laser drives the ground state $\left|g\right\rangle $ and the Rydberg state $\left|r_{1}\right\rangle $. For experiment, three lasers are used to excite the atoms, as shown in the dashed box. A strong MFC field is set as a local oscillator (LO) field (multicolor), and a signal field (blue) couples the two Rydberg states $\left|r_{1}\right\rangle $ and $\left|r_{2}\right\rangle $.
(b) Signal beat (blue) with two MFCs (red). The beat notes frequency  are determined by frequencies of the signal field and its nearest-neighbor MFC comb line. From the sum and difference of the $f_{b}\pm f_{b^{\prime}}$, we calculate the mode-order number $N$ of the nearest-neighbor MFC comb line. The absolute frequency of the signal field is obtained from the mode-order number and its offset from its nearest MFC frequency component.
(c) Experimental setup. A 852-nm probe laser, a 1470-nm dressing laser, and a 779-nm coupling laser are used to excite and probe the Rydberg state. The MFC and signal fields are combined through a resistance power divider (RPD), and transmitted to the vapor cell by a horn antenna. The beat note signal is recorded and analyzed via an oscilloscope (OSC) or a spectrum analyzer (SA).}
\label{fig1}
\end{figure*}

The energy level is depicted in Fig.~\ref{fig}(a), and consists of a ground state $\left|g\right\rangle $ and two Rydberg states $\left|r_{1}\right\rangle $ and $\left|r_{2}\right\rangle $. A laser resonantly couples the ground state $\left|g\right\rangle $ and the Rydberg state $\left|r_{1}\right\rangle $ with Rabi frequency $\Omega$. An MFC field $E_{\mathrm{MFC}}$ and a signal field $E_{\mathrm{s}}$, with Rabi frequencies $\sum_{i}\Omega_{\mathrm{LO_{i}}}$ and $\Omega_{\mathrm{s}}$, drive the Rydberg RF transitions $\left|r_{1}\right\rangle \leftrightarrow\left|r_{2}\right\rangle $. Then these two MW fields are mixed via Rydberg atoms \cite{simons2019rydberg,jing2020atomic} through the interaction between the microwave fields and Rydberg atoms, resulting in several beat notes between the signal and MFC comb lines which is read from the transmission spectrum. And this can be regarded as the beat effect of the transition probability \cite{PhysRevA.104.053101}. As the frequency of the microwave field is 
much higher comparing with the difference-frequency part of the beat note, the sum-frequency part of the beat note is ignored, and the difference-frequency part is reserved. The strongest beat is the one between the signal and its nearest-neighbor MFC line. In addition, at least two MFCs with different mode spaces are used to get the absolute frequency of the signal, since the frequency of the beat note between one MFC field and the signal field only contains the information on the relative frequency of the signal to its nearest-neighbor MFC comb line \cite{2003A}.

The total field in the atoms is written as \cite{simons2019rydberg}
\begin{align}
\begin{split}
E_{\text{atom}}&=\sqrt{|E_{\mathrm{L}}|^{2}+|E_{\mathrm{s}}|^{2}+2|E_{\mathrm{L}}||E_{\mathrm{s}}|\cos(\Delta\omega t+\Delta\phi)}\\ &\approx  |E_{\mathrm{L}}|+|E_{\mathrm{s}}|\cos(\Delta\omega t+\Delta\phi)
\end{split}
\label{eq}
\end{align}
here, the single local oscillator (LO) field term $E_{\mathrm{L}}$ is replaced by an MFC field $E_{\mathrm{MFC}}=\sum_{i}E_{\mathrm{LO}_{i}}\cos(\omega_{i}t+\phi_{i})$, in which $E_{\mathrm{LO}_{i}}=E_{\mathrm{LO}}$ corresponds to the strength of the $i^{th}$ frequency comb line. Because of the limited bandwidth of the system \cite{meyer2018digital}, we set the MFC repetition rate $f_{\mathrm{r}}$ larger than the instantaneous bandwidth of the system. The field sensed by the Rydberg atoms can be expressed as
\begin{equation}
E_{\mathrm{atom}}\approx\sqrt{N_{L}}|E_{\mathrm{LO}}|+\frac{1}{\sqrt{N_{L}}}|E_{\mathrm{s}}|\cos(\Delta\omega_{k}t+\Delta\phi_{k})
\label{eq1}
\end{equation}
where $N_{L}$ represents the number of MFC comb lines. The subscript $k$ represents the number of the MFC line nearest to the signal field. When $E_{\mathrm{s}} \ll E_{\mathrm{LO}}$, the beat term between MFC comb lines, $2|E_{\mathrm{LO}_{i,i\neq k}}||E_{\mathrm{s}}|\cos(\Delta\omega_{i} t+\Delta\phi_{i})$,
is small compared with the square of the LO electric field strength $|E_{\mathrm{LO}}|^{2}$.
According to Eq.~\ref{eq1}, the signal amplitude and phase are extracted from its beat note
with the nearest-neighbor MFC comb line.
As shown in Fig.~\ref{fig1}(b), two MFCs are used (denoted as $E_{\mathrm{MFC_{1}}}$ and $E_\mathrm{{MFC_{2}}}$) to measure the absolute frequency of the signal field $E_{\mathrm{s}}$. For a proof-of-principle study, we turn on one of the MFCs at a different time. We set these two MFCs with the same offset frequency $f_{\mathrm{offset}}$ and with different repetition rates. The difference between the repetition rates is marked as $\delta f_{\mathrm{r}}$, and it satisfies $\delta f_{\mathrm{r}}=f_{\mathrm{r}}-f_{\mathrm{r}^{\prime}}\ll f_{\mathrm{r}}$. The exact-fraction method in Ref.~\cite{2003A} is employed to calculate the frequency of the signal  $f_{\mathrm{s}}$. Its expression is as follows:
\begin{equation}
f_{\mathrm{s}}=f_{\mathrm{offset}}+N f_{\mathrm{r}}\pm f_{\mathrm{b}}\,
\label{eq3}
\end{equation}

\begin{figure}[btp]
\includegraphics[width=1\linewidth]{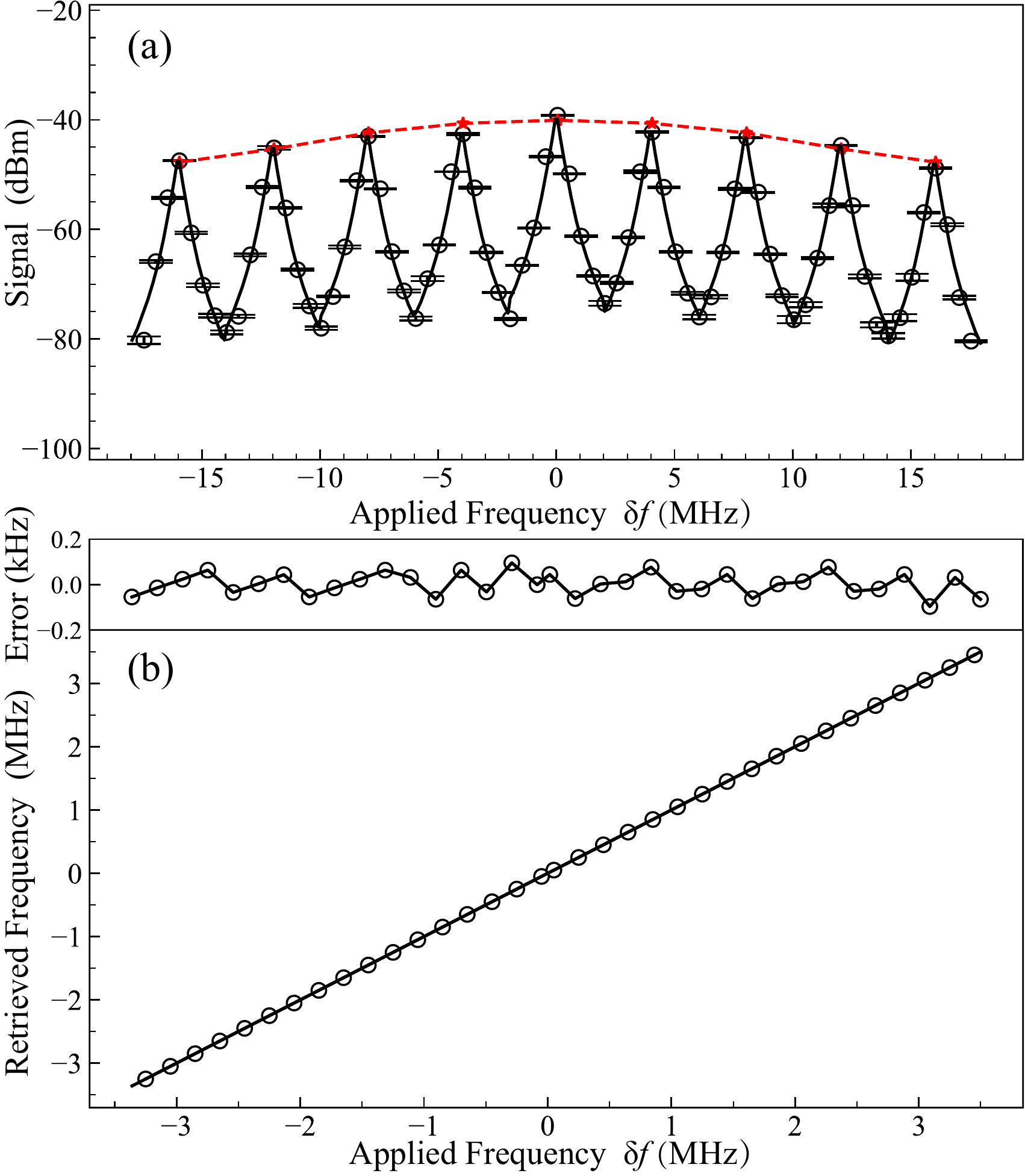}\caption{(a) The power of the beat note versus the signal field frequency offset $\delta f$. The experimental data (black circles), the semi-steady-state theoretical result (red dashed line), and the time-dependent calculation results (black solid line) are shown. The beat note is recorded with a frequency separation of 500\,kHz. Here, a 9-bins MFC field is used. (b) The retrieved frequency versus the signal field frequency. Here, two 25-bins MFCs are used. The subfigure shows the frequency error $\sim$ 100\,Hz.}
\label{fig2}
\end{figure}

The mode-order number of the signal is calculated according to the sum/difference of the two beat-note frequencies  $f_{b}$, $f_{b^{\prime}}$. The mode-order number for the MFC1 and MFC2 is marked as $N_{1}$ and $N_{2}$ (start from 0). And the two MFCs both have $N_{c}$ comb lines. Assume the repetition rates satisfy $f_{r} >f_{{r}^{\prime}}$. The calculation of the mode-order number $N$ are divided into 2 cases. 
Case 1: $N_{2}-N_{1}=1$ When $f_{r}-(f_{b}+{f_{b}}^{\prime})\leq(N_{c}-2)\delta f_{r}$, the mode-order number  of the MFC2 is $(f_{r}-f_{b}-{f_{b}}^{\prime})/\delta f_{r}$-1. Case 2: $N_{1}=N_{2}$. If the condition in case 1 is not satisfied, the mode-order of the MFC2 is given by $N_{2}=(f_{b}+{f_{b}}^{\prime})/\delta f_{r}$ or $N_{2}=(|f_{b}-{f_{b}}^{\prime}|)/\delta f_{r}$. And if $f_{b}+{f_{b}}^{\prime}> N_{c}$, $N_{2}=(|f_{b}-{f_{b}}^{\prime}|)/\delta f_{r}$.
The calculation of the mode-order number with part of the experiment data measured in Fig.~\ref{fig2}(b) is given by the table \ref{tab1}. 
At last, the signal frequency can be calculated by through the Eq.~\ref{eq3}.

\begin{table*}
\caption{The mode-order number calculation}
\begin{tabular}{|l|l|l|}
\hline  & Case1 ($N_{1} - N_{2} = 1$) & Case2 ($N_{1} = N_{2}$) \\
\hline Beat1 Frequency- $\mathrm{f}_{b}$ & $132.104\,\mathrm{kHz}$ & $62.92\,\mathrm{kHz}$ \\
\hline Beat2 Frequency- $\mathrm{f}_{b}^{\prime}$ & $130.019\,\mathrm{kHz}$ & $49.955\,\mathrm{kHz}$ \\
\hline Mode-order number & $\mathrm{~N}_{2}=\left(280-\mathrm{f}_{b}-\mathrm{f}_{b}^{\prime}\right) / 1-1=16.877 \approx 17$ & $\mathrm{N}_{2}=(\mathrm{f}_{b}-\mathrm{f}_{b}^{\prime}) / 1=12.965 \approx 13$ \\ \hline
\end{tabular}
\label{tab1}
\end{table*}

The mode-order number $N$ is often not an integer because of the finite measurement precision of the beat frequencies $f_{\mathrm{b}}$ and $f_{\mathrm{b^{\prime}}}$. To correctly estimate $N$, the resolution bandwidth of the data acquisition device (e.g., the resolution bandwidth of a spectrum analyzer) should be set larger than $\delta f_{\mathrm{r}}$. The value $\delta f_{\mathrm{r}}$ satisfies the formula $N \delta f_{\mathrm{r}} <f_{\mathrm{r}}/2$. This guarantees that the mode-order numbers of the signal with comb 1 and comb 2 have a difference less than 1. As the beat note with the nearest-neighbor comb line is the strongest one within a range of $f_{\mathrm{r}}/2$, we only monitor the signal in this range. As a proof-of-principle experiment, we turn on the two MFCs at different times and keep the amplitude of the signal unchanged to demonstrate the frequency measurement [In practice, using two vapor cells applied with two different MFCs could be an effective method for real-time frequency measurement].

We use three infrared lasers \cite{Carr:12,PhysRevA.100.063427} to excite the Cs atoms from the ground state $\left|g\right\rangle $ to the Rydberg state $\left|r_{1}\right\rangle $. The laser beams consist of probe, dressing, and coupling lasers. The probe laser is resonant with the transition $6S_{1\text{/2}}, F=4\rightarrow6P_{3/2}, F'=5$($\left|1\right\rangle \rightarrow \left|2\right\rangle$). And the dressing laser couples the transition $6P_{3/2}, F'=5\rightarrow7S_{1/2}, F''=4$ ($\left|2\right\rangle \rightarrow \left|3\right\rangle$) through two-photon spectrum. The probe beam is split into two beams and collected by a balanced photodetector. Both of the coupling and dressing beams counter-propagate with the probe beam. We use a differential detection scheme to measure the transmission spectrum. The MFC and signal fields interact with Rydberg atoms simultaneously, and the transmission spectrum displays an interference beating via the mixing process in Rydberg atoms. This beat note is extracted from the transmission spectra via a low-pass filter [Fig.~\ref{fig1}(c)]. We use a spectrum analyzer to collect the frequency and the amplitude of this signal.

\begin{figure*}
	\includegraphics[width=1\linewidth]{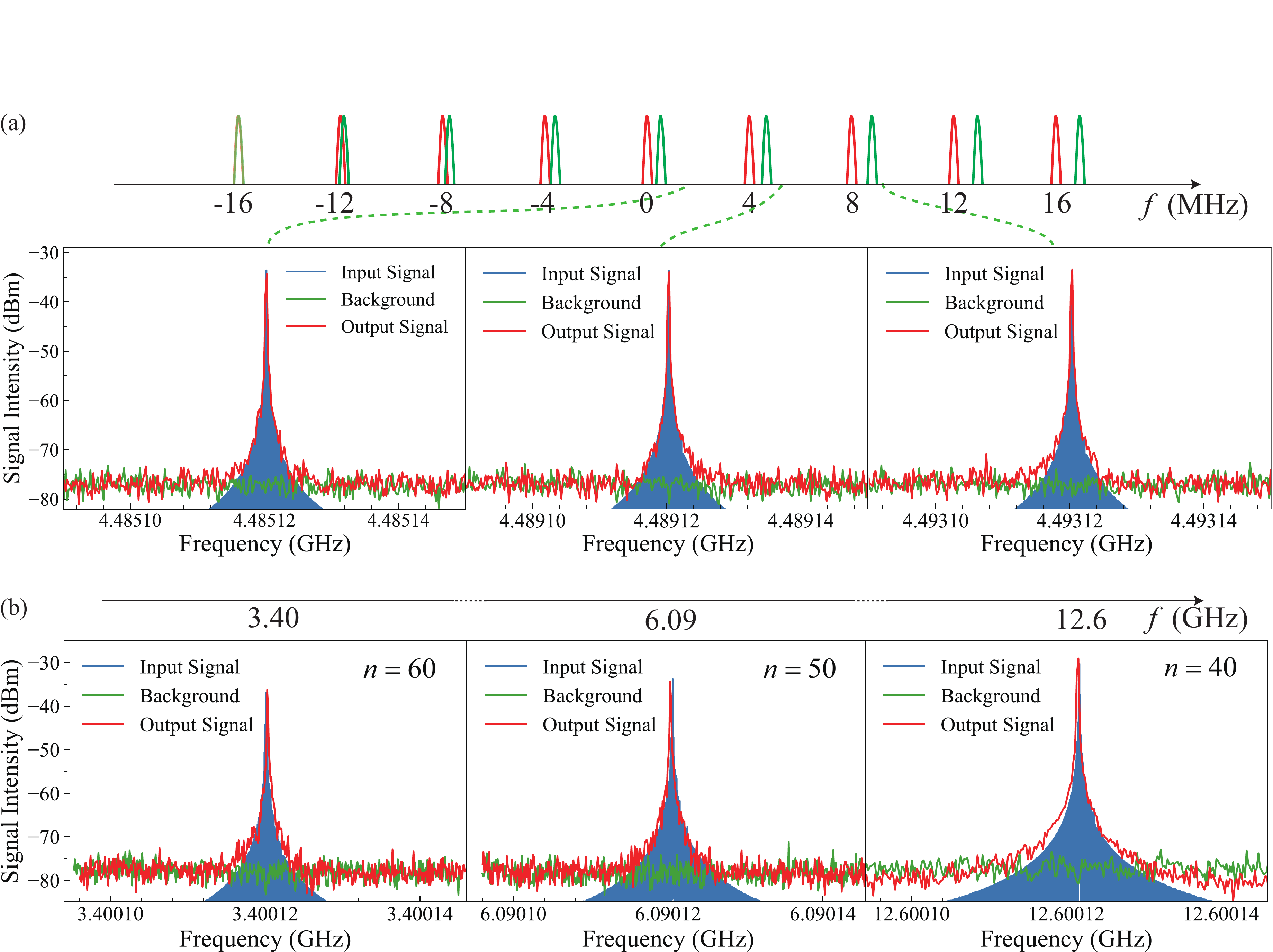}\caption{(a)The received 1-kHz bandwidth signal spectrum near different comb bins. (b)The measured frequency spectrum for a signal with 1\,kHz bandwidth: the amplitude of the frequency spectrum of the output beat notes using MFCs with central frequencies of 3.40\, GHz, 6.09\,GHz and 12.6\,GHz. The used Rydberg state is $60P_{3/2}$, $50P_{3/2}$, $40P_{3/2}$, respectively.}
	\label{fig3}
\end{figure*}

\section{performance for frequency measurement}
To demonstrate the wide response range of the Rydberg MFC spectrometer, we measure the amplitude of the output beat note versus the signal field frequency $f_{\mathrm{s}}$. By fixing the power of the signal MW field to $\sim$ -30\,dBm, we change the frequency offset of the signal $\delta f$ with respect to the Rydberg RF transition 55$P_{3/2} \longleftrightarrow 54D_{5/2}$ (with 4.485\,GHz) from -18\,MHz to +18\,MHz. The amplitude of the output beat note is shown in Fig.~\ref{fig2}(a). There are nine peaks, which correspond to the number of applied MFC comb lines. Each peak reflects the beat note response between the individual comb line component and the signal. And the peak profile is related to the finite instantaneous band of the system [the 3-dB instantaneous band $\delta{w}$ is 300\,kHz for our system], which is determined by the time for the system to reach the steady state \cite{meyer2018digital}. This finite band can be described by optical Bloch equations, which should consider the time-dependent response of the atomic system (see the further analysis and estimation of the instantaneous bandwidth in Appendix~\ref{append:insband} and Appendix~\ref{app:theorycal}). The black solid curve in Fig.~\ref{fig2}(a) shows the theoretical result, which is normalized to the maximum amplitude of the signal at each peak. The instantaneous bandwidth $\delta{w}$ is small compared with half of the repetition rate $f_{\mathrm{r}}$ of the MFC, e.g., 2\,MHz, the observed power of the beat note decreases rapidly when the frequency of the signal gets away from its nearest-neighbor MFC comb line. Furthermore, the maximum response
appears when the signal is close to the MFC comb line. With increasing signal frequency offset $\delta{f}$ from the resonance, the beat note response is reduced. The theoretical calculation is the red dashed line. This is modeled by numerically solving the semi-steady state of the master equation. The response when applying an MFC with 25 comb lines is illustrated in Fig~.\ref{fig6} of Appendix~\ref{app:mfcset} to show the maximum response of the system.

In Fig.~\ref{fig2}(b), we show the retrieved frequency versus the signal frequency when two MFCs (25 comb lines) with a covering range of $\sim$ 7\,MHz are applied. The repetition rates of these two MFCs, $f_{\mathrm{r}}$ and $f_{\mathrm{r^{\prime}}}$, are 280\,kHz and 279\,kHz, respectively. The difference in the repetition rate is $\delta f_{\mathrm{r}}=f_{\mathrm{r}}-f_{\mathrm{r^{\prime}}}=1$\,kHz. We set the resolution bandwidth of the spectrum analyzer, $\Delta_{\mathrm{RBW}}$, as 30\,Hz (less than $\delta f_{\mathrm{r}}$) to extract the mode-order number $N$. Thus, the deviation between the input and the retrieved signal is determined by the resolution of the measurement device, as shown in Fig.~\ref{fig2}(b). In our experiment, the resolution bandwidth of the spectrum analyzer, $\Delta_{\mathrm{RBW}}$, is set as 10\,Hz. This protocol has advantages over the method of scanning the frequency of a single LO field \cite{jing2020atomic}, since there is no extra requirement to adjust the LO frequency to capture the signal in the limited bandwidth. This realizes real-time signal detection in a wide range.

\begin{figure}
\includegraphics[width=1\linewidth]{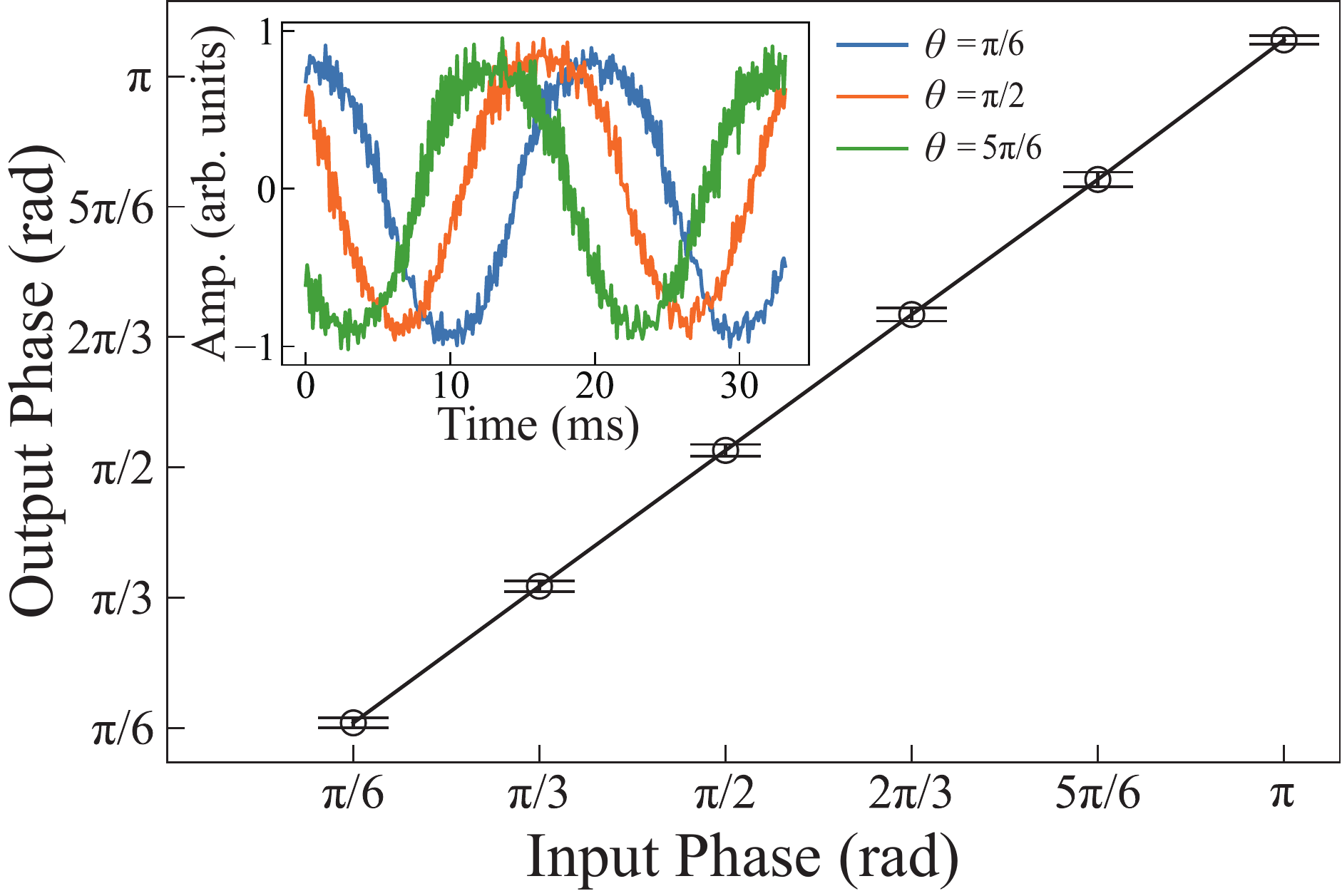}\caption{The output phase versus the input phase. The inset is the  recorded transmission signal. The extra noise in the transmission signal is mainly from the beat notes between the different MFC comb lines. The power of the signal field is -20\,dBm.}
\label{fig4}
\end{figure}

\section{Receiving an 1\,kHz bandwidth signal}
In practice, the input MW field often has a certain bandwidth and different central frequencies, such as the MW field emitted from free radicals \cite{dousmanis1955microwave}. We demonstrate the ability of the system to measure the frequency of this kind of signal. In Fig.~\ref{fig3}(a), the signal frequency spectrum received by different comb bin is shown.  The input signal frequency spectrum and the output beat-note frequency spectrum is shown in Fig.~\ref{fig3}(b), in which the selected different principal quantum numbers $n$ are used to detect different frequencies. For the guide of the eyes, the power of the input signal are all shifted with -10 dBm, and the profiles of output signal are in good agreement with the input signal. The absolute frequency of the input signal is calculated through the frequency difference of the beat note. In this process, the bandwidth of the signal (1\,kHz) is less than the repetition rate of the MFC. 

\section{phase recognition}
The MFC sensing method allows recognition of the signal phase. We measure the output phase by changing the phase of the input signal field. The phase information is extracted from the sinusoidal fit coefficient in the formula $y=A\mathrm{sin}(\omega t+\phi)$. We set the MFC field with three comb lines and a frequency repetition rate of 300\,kHz, and set the signal frequency offset $f_{\mathrm{r}}$ as 300.05 kHz. The output phase versus the input signal phase is shown in Fig.~\ref{fig4}. The inset of Fig.~\ref{fig4} corresponds to the transmission signal with an input phase of $\pi/6,\pi/2,5\pi/6$ which shows that the transmission signal is phase-sensitive to the input signal. The linear regression coefficient is 0.98 for the input and output phase. And the reduced chi-square is 0.63 for this regression showing a linear dependency function, in which the small phase error is from the fluctuation of system.

\section{Discussion}
Compared with the heterodyne method \cite{jing2020atomic} of addressing a single LO field, the demonstrated MFC method can measure the absolute frequency of a signal. This could promote applications in radar and communications because the heterodyne method can only detect the offset of the signal frequency $\left|\delta f\right|$ and has difficultly distinguishing the sign of the frequency, which reveals an object's direction of motion. As limited by the evolution time to reach the steady state, the instantaneous bandwidth of the Rydberg MW sensor is typically less than 10 MHz in the traditional methods \cite{meyer2018digital}, while the MFC method can have an instantaneous working bandwidth of hundreds of MHz considering the strong off-resonant response of the Rydberg atoms \cite{Waveguide2021}. In addition, we considered the master equation in a simple way to describe the results. Since the MFC fields consist of multiple frequency lines, a more precise method such as a Floquet calculation would be more effective.

The MFC spectrometer based on Rydberg atoms has many potential advantages, for example: 1) There is no need to change the size of the Rydberg receiver as the classical antenna should have its size changed to match MW electric fields with different frequencies; 2) Rydberg atoms could be used to achieve high sensitivity via the mixing method \cite{jing2020atomic}. A series of optimized operations could improve the performance of this Rydberg MFC spectrometer, include locking the lasers to an ultra-stable cavity, reducing the system noise by using homodyne detection, and reducing the coupling loss of MW electric fields by a waveguide \cite{Waveguide2021}.

\section{Conclusion}
In summary, we have demonstrated an absolute frequency measurement through a Rydberg MFC spectrometer. The instant detection range can be further increased by improving the MFC bandwidth, so there is still room for improvement. The high-resolution spectroscopy of the Rydberg atoms allows precise detection of the electric fields and the Rydberg state \cite{PhysRevLett.82.1831,PhysRevLett.98.113003}. Further study on various MW dressing protocols may improve the sensitivity, bandwidth, and precision in Rydberg atomic sensing \cite{jing2020atomic,meinel2021heterodyne,Bason_2010}. Rydberg atoms also possess high sensitivity in the terahertz frequency range \cite{PhysRevX.10.011027,Chen:22}. Thus combining Rydberg atoms with a terahertz frequency comb \cite{PhysRevX.4.021006,yasui2006terahertz} also permits precise, real-time, and coherent measurement of terahertz signals in communication and metrology.

\begin{acknowledgments}
We acknowledge funding from the National Key R\&D Program of China (Grant No. 2017YFA0304800), National Natural Science Foundation of China (Grant Nos. U20A20218, 61525504, and 61435011), Youth Innovation Promotion Association of the Chinese Academy of Sciences (Grant No. 2018490) and the Major Science and Technology Projects in Anhui Province (Grant No. 202203a13010001).
\end{acknowledgments}

\appendix
\section{MFCs setup}
\label{app:mfcset}
\begin{figure}[htp!]
	\includegraphics[width=0.85\linewidth]{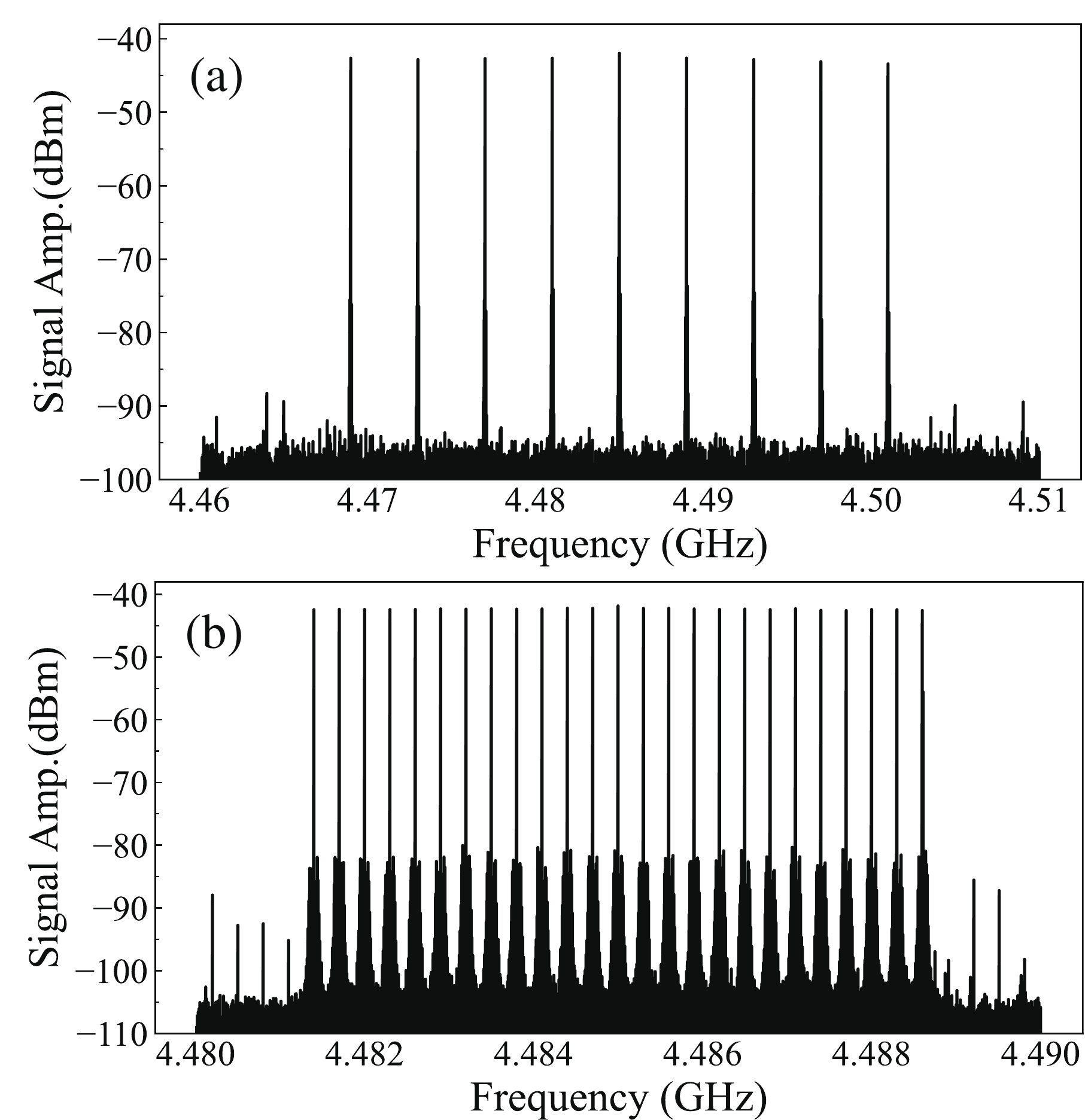}\caption{The frequency spectrum of the MFCs used in Fig.~\ref{fig2} are plotted. (a) The frequency spectrum of the MFC used in Fig.~\ref{fig2}(a). (b) Frequency spectrum for one of the MFCs used in Fig.~\ref{fig2}(b).}
	\label{fig5}
\end{figure}
\begin{figure}[htp!]
	\includegraphics[width=0.85\linewidth]{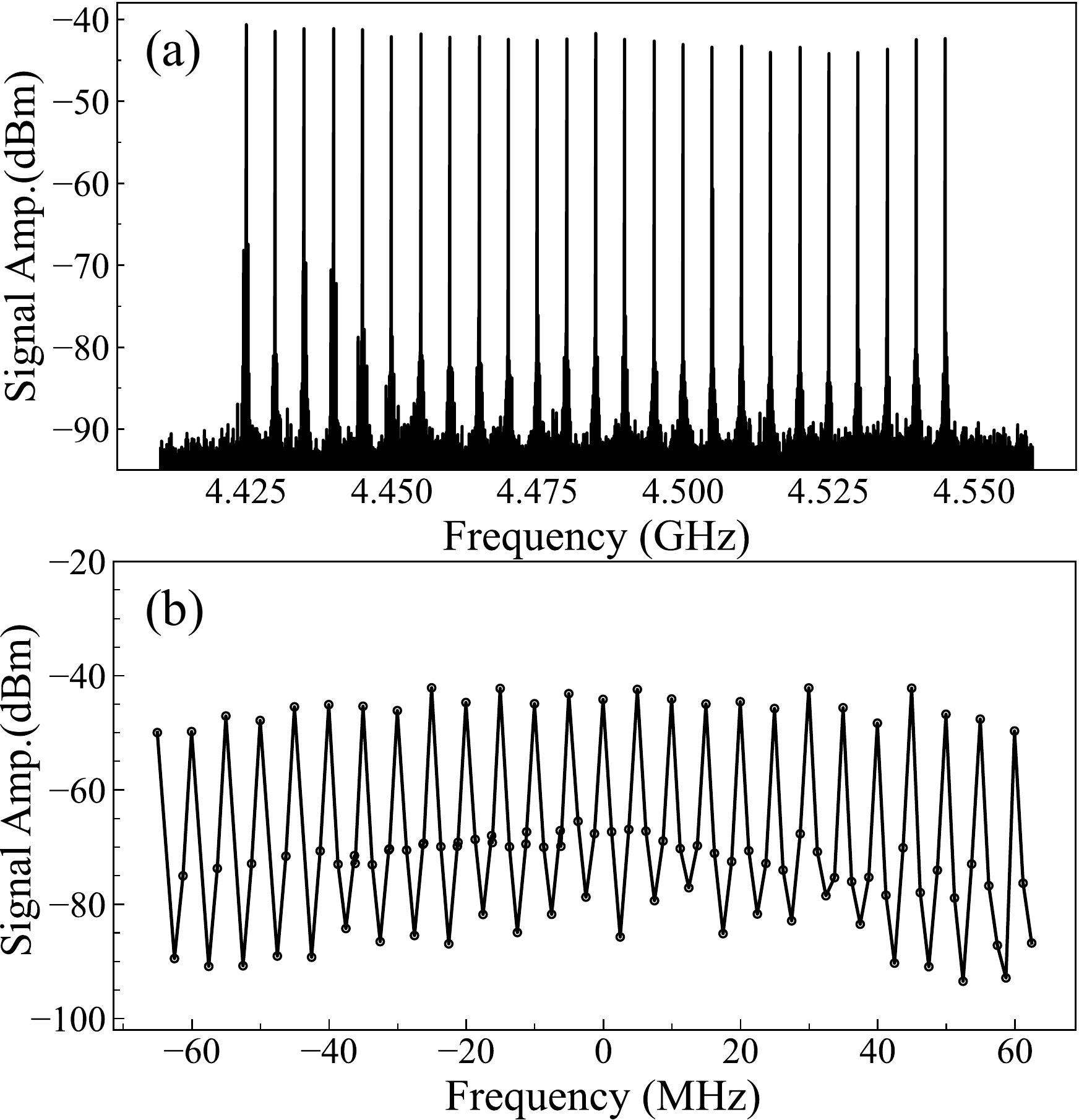}\caption{(a) The spectrum of the MFC used to demonstrate the system maximum response range (125\,MHz) is shown. (b) The power of the beat note response versus the signal frequency when applying the MFC covering range of 125\,MHz. Here, we set the signals with a constant power of $\sim$ -20\,dBm.}
	\label{fig6}
\end{figure}

\begin{figure}[htp]
 \centering
	\includegraphics[width=0.85\linewidth]{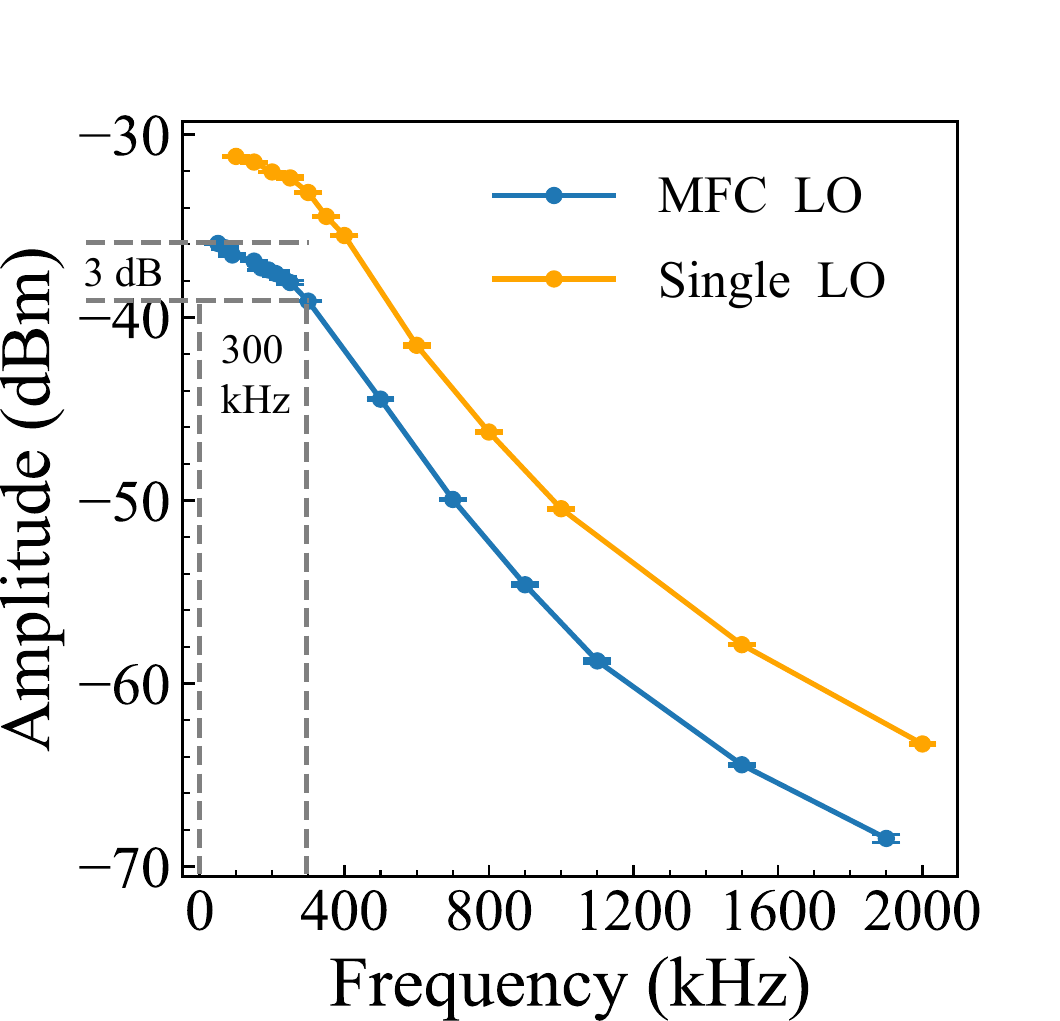}\caption{The power of the beat note as a function of its frequency. The  3-dB instantaneous bandwidth is about 300\,kHz indicated by the dashed line.}
	\label{fig7}
\end{figure}

As we know, a frequency comb consists of phase-stabilized narrow frequency lines with equidistant mode spacing (repetition frequency). Here, we use the multi-tone modulation of the vector signal generator (1465V Ceyear) to simulate a MFC signal. The central frequency of each MFC used in our experiment is set near 4.485\,GHz to maximize the beat note response to the signal field. The repetition rates of the MFC used in Fig.~\ref{fig2} (a) and (b) are 4\,MHz and 280\,kHz, respectively. The frequency spectrum of the MFC used in our experiment is acquired through the spectrum analyser (4024F Ceyear). The MFC frequency spectra are shown in Fig.~\ref{fig5}. In the MFC frequency spectra, the main comb line components dominate, so the effects of other comb line components are weak. Besides, we also plot the power of the output beat note versus the applied signal frequency when an MFC with a range of 125\,MHz is applied in Fig.~\ref{fig6} (b). Finally, the maximum comb span of the MFC is limited by the maximum clock frequency of the signal generator (about 125\,MHz). Other MFC generation methods are supposed to overcome this limitation to achieve a wider measurement range.

\section{Instantaneous bandwidth}
The instantaneous bandwidth in this Rydberg atom MW receiving system refers to the first frequency of the (beat note power) response reaching $50\%$ (-3dB) of its maximum value. We record the system response power versus the frequency of the beat note to measure this bandwidth. The frequency response of the MFC scheme is measured within a span of the MFC repetition rate, so the signal MW field only beat with one of the MFC comb line. In Fig.~\ref{fig7}, the instantaneous bandwidths are almost the same for the MFC LO and the single LO scheme, and the frequency response character (the profile of the response curve) is similar. This bandwidth is dependent on the relaxation time for the atoms to reach a steady state. And for our experiment system, it is about 300\,kHz. In the rubidium Rydberg atoms system, the bandwidth is several MHz \cite{simons2019rydberg} which would be more helpful to achieve a flat beat note response.
\label{append:insband}
\section{Theoretical calculation}
The dynamics of this system is described using the Lindblad equation \citep{jing2020atomic,PhysRevA.100.063427}
\begin{equation}
\dot{\hat{\rho}}=\frac{i}{\hbar}[\hat{\rho},\hat{H}]+L(\hat{\rho})
\label{eq:master}
\end{equation}
There are two regimes of beat-note response in the Fig.~\ref{fig2}(a): (i) When the beat-note frequency $f_{b}$ (e.g., 50\,kHz) is less than the system instant bandwidth ($\sim$ 300\,kHz), the power of the beat note is calculated through solving the steady-state Lindblad eqution which corresponds to the maximum signal at the top of each peak (Red) in Fig.~\ref{fig2}(a); (ii) When $f_{b}>$ 300\,kHz, we model the beat-note power reduction compared with its maximum value by a time-dependent optical bloch equation and the numerical results are shown using solid lines in Fig.~\ref{fig2}(a).

For the case $f_{b}<$ 300\,kHz, the system Hamiltonian under the bare atom basis $\left|1\right\rangle$, $\left|2\right\rangle$, $\left|3\right\rangle$, $\left|4\right\rangle$, and $\left|5\right\rangle$ is
\begin{equation}
\hat{H}=\left(\begin{array}{ccccc}
0 & \Omega_{p}/2 & 0 & 0 & 0\\
\Omega_{p}/2 & \Delta_{1} & \Omega_{d}/2 & 0 & 0\\
0 & \Omega_{d}/2 & \Delta_{2} & \Omega_{c}/2 & 0\\
0 & 0 & \Omega_{c}/2 & \Delta_{3} & \Omega_{MW}(t)/2\\
0 & 0 & 0 & \Omega_{MW}(t)/2 & \Delta_{4}
\end{array}\right)    
\end{equation}
And the Lindblad term in the above Eq.~\ref{eq:master} is:
\begin{widetext}
\begin{equation}
\ensuremath{L(\hat{\rho})=\left(\begin{array}{ccccc}
\Gamma_{e}\rho_{22} & -\frac{1}{2}\Gamma_{e}\rho_{12} & -\frac{1}{2}\Gamma_{d}\rho_{13} & -\frac{1}{2}\Gamma_{r1}\rho_{14} & -\frac{1}{2}\Gamma_{r2}\rho_{15}\\
-\frac{1}{2}\Gamma_{e}\rho_{21} & -\Gamma_{e}\rho_{22}+\Gamma_{d}\rho_{33}+\Gamma_{r2}\rho_{55} & -\frac{1}{2}\left(\Gamma_{e}+\Gamma_{d}\right)\rho_{23} & -\frac{1}{2}\left(\Gamma_{e}+\Gamma_{r1}\right)\rho_{24} & -\frac{1}{2}\left(\Gamma_{e}+\Gamma_{r2}\right)\rho_{25}\\
-\frac{1}{2}\Gamma_{d}\rho_{31} & -\frac{1}{2}\left(\Gamma_{e}+\Gamma_{d}\right)\rho_{32} & -\Gamma_{d}\rho_{33}+\Gamma_\mathrm{{r1}}\rho_{44} & -\frac{1}{2}\left(\Gamma_{d}+\Gamma_{r1}\right)\rho_{34} & -\frac{1}{2}\left(\Gamma_\mathrm{{d}}+\Gamma_{r2}\right)\rho_{35}\\
-\frac{1}{2}\Gamma_{r1}\rho_{41} & -\frac{1}{2}\left(\Gamma_\mathrm{{e}}+\Gamma_{r1}\right)\rho_{42} & -\frac{1}{2}\left(\Gamma_{d}+\Gamma_{r1}\right)\rho_{43} & -\Gamma_{r1}\rho_{44} & -\frac{1}{2}\left(\Gamma_\mathrm{{r1}}+\Gamma_{r2}\right)\rho_{45}\\
-\frac{1}{2}\Gamma_{r2}\rho_{51} & -\frac{1}{2}\left(\Gamma_\mathrm{{e}}+\Gamma_{r2}\right)\rho_{52} & -\frac{1}{2}\left(\Gamma_{d}+\Gamma_\mathrm{{r2}}\right)\rho_{53} & -\frac{1}{2}\left(\Gamma_{r1}+\Gamma_{r2}\right)\rho_{54} & -\Gamma_{r2}\rho_{\mathrm{55}}
\end{array}\right)}
\end{equation}
\end{widetext}
where, $\Omega_{p}$, $\Omega_{d}$, $\Omega_{c}$, and $\Omega_{MW}(t)$ refer to the Rabi frequencies of the probe, dressing, coupling laser, and the MW field, respectively, which are defined as $\Omega_{i}=\mu_{i}E_{i}/\hbar$. $\mu_{i}$ and $E_{i}$ is the corresponding dipole moment and the intensity of the electric field. $\Delta_{p}$, $\Delta_{c}$, $\Delta_{d}$, and $\Delta_{MW}$ present the zero-velocity atom-field-detunings of all the fields above, respectively. $\Gamma_{r1}$, $\Gamma_{r2}$, $\Gamma_{e}$, and $\Gamma_{d}$ represent the decay rates of the two Rydberg states and two intermediate states, respectively.

\begin{figure*}[htp!]
	\includegraphics[width=1\textwidth]{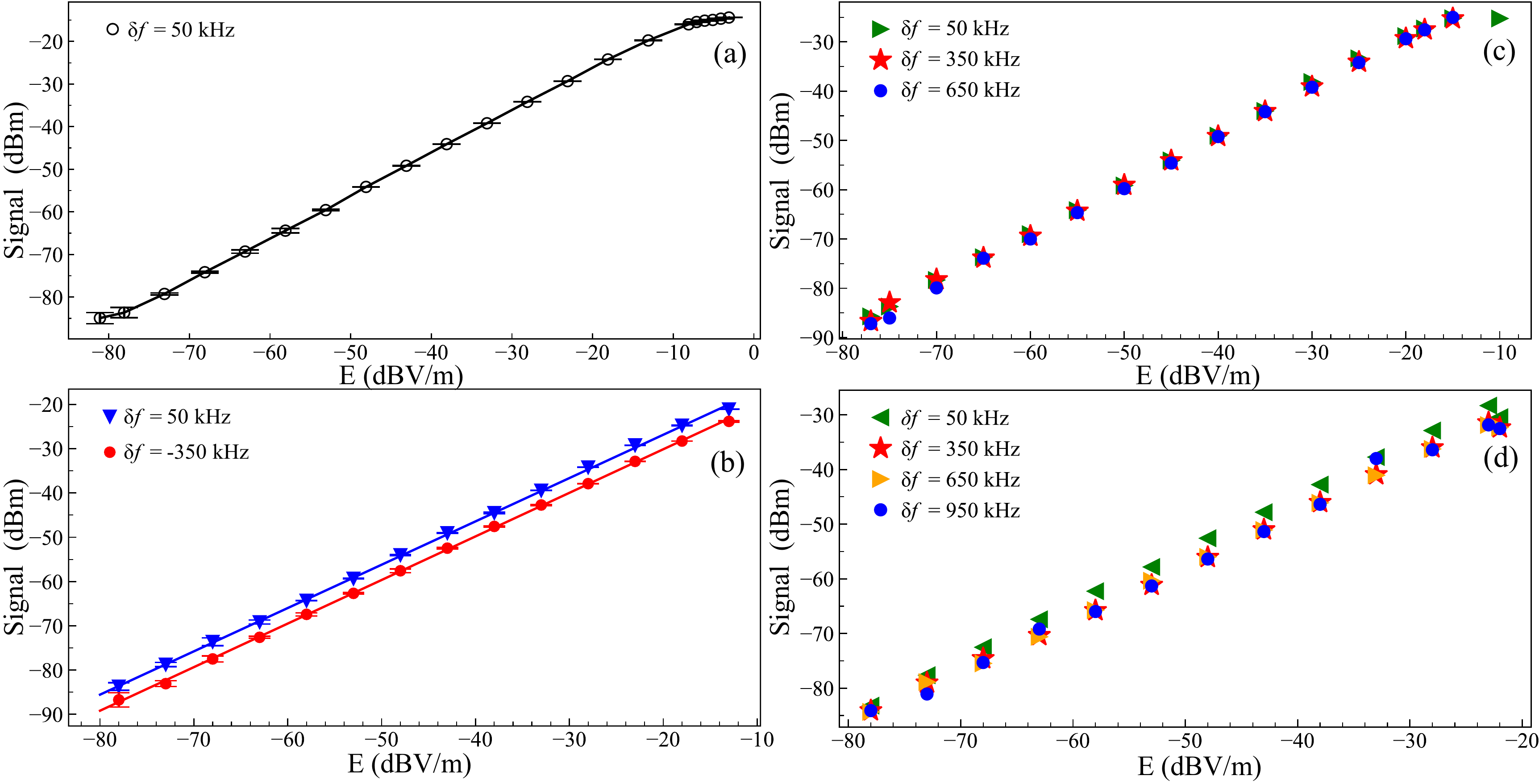}\caption {The measured dynamic range when applying the MFC field with 3, 5, 7 comb lines. The MFCs' frequency repetition rate $f_{r}$ is 300\,kHz. The frequency offset is set as $\delta f = 50$ kHz (green), $350$ kHz (red), $650$ kHz (orange) and 950 kHz (blue), respectively.}
	\label{fig8}
\end{figure*}

In our experiment, we use the MFC field as a strong LO field. The intensity of the MFC field is $E_{\mathrm{MFC}}=\sum_{i}E_{LO_{i}}$, in which $E_{LO_{i}}$ is the intensity of the $i^{\mathrm{th}}$ frequency comb line. We assume that each of the MFC line has the same strength $|E_{LO_{1}}|=|E_{LO_{2}}|=\cdots=|E_{LO_{i}}|=|E_{LO}|$. When the beat note of the signal is small compared with the total LO field power $|E_{\mathrm{s}}| \ll |E_{LO}|$, we have $2\sum_{i}|E_{LO_{i}}||E_{\mathrm{s}}|\cos(\Delta\omega_{i} t+\Delta\phi_{i}) \ll N_{L}|E_{LO}|^{2}$. The total field sensed by the Rydberg atoms is expressed as:
\begin{align}
E_{\mathrm{atom}}\approx\sqrt{N_{L}}|E_{LO}|\sqrt{\begin{array}{c}
1+2\frac{\sum_{i}|E_{\mathrm{s}}|\cos(\Delta\omega_{i}t+\Delta\phi_{i})}{N_{L}|E_{LO}|}\\
+\frac{\sum_{ij}\cos(\Delta\omega_{ij}t+\Delta\phi_{ij})}{N_{L}}
\end{array}}
\end{align}
Considering the instant bandwidth of the system and we set the MFC reptition rate (e.g., $\sim$ 4\,MHz) larger than the instant bandwidth of the system ($\sim$ 300\,kHz), the response of the system to the high-frequency interference terms between different MFC lines $\sum_{ij}^{i\neq j}|E_{LO_{i}}||E_{LO_{j}}|\cos(\Delta\omega_{ij}t+\Delta\phi_{ij})$ is negligible, and the high-order nonlinear response is assumed when $E_{LO_{i}}$ is not strong (less than -10 dBm). Thus, we consider that the $k^{th}$ MFC line is the nearest neighborhood to the signal field. 
\begin{align}
E_{\mathrm{atom}} \approx \sqrt{N_{L}}|E_{LO}|+\frac{1}{\sqrt{N_{L}}}|E_{\mathrm{s}}|\cos(\Delta\omega_{k}t+\Delta\phi_{k})
\label{eq4}
\end{align}
 Thus, the amplitude and phase of the signal are encoded in its beat note with the nearest-neighbor MFC comb line which is read from the Rydberg nonlinear spectra transmission signal. 

In the interaction picture, the detunings  $\Delta_{1}=\Delta_{p}$, $\Delta_{2}=\Delta_{p}+\Delta_{c}$, $\Delta_{3}=\Delta_{p}+\Delta_{c}+\Delta_{d}$ and $\Delta_{4}=\Delta_{p}+\Delta_{c}+\Delta_{d}-\Delta_{MW}$.   
The probe transmission is expressed as:
\begin{equation}
    P=P_{0}e^{(-2\pi L \mathrm{Im}[\chi]/\lambda_{p})}
\end{equation}
where $P$ is the probe laser transmission, $P_{0}$ is the the laser power being input in the vapor cell, $\chi$ is the susceptibility of the probe laser  $\chi=\frac{2N_{0}\mathcal{D}_{12}}{E_{p}\epsilon_{0}}\rho_{21}$, L is the path of the laser in the cell, $N_{0}$ is the total atom  density in the vapor cell, $\mathcal{\mu}_{12}$ is the dipole moment associated with the ground state transition, $E_{p}$ denotes electric-field amplitude of the probe laser, and $\epsilon_{0}$ is the vacuum electric permittivity. For maximum measurement sensitivity, we record the Rydberg MFC spectrum on resonance with the condition $\Delta_{p}=\Delta_{d}=\Delta_{c}=0$. The instantaneous steady-state solution of the density matrix element  $\rho_{21}$ is calculated through the Eq.~\ref{eq:master}
{\label{theory2}

The frequency response curve shown in Fig.~\ref{fig2}(a) is modeled through the instant bandwidth model based on the theory \cite{meyer2018digital,PhysRevA.69.063801,PhysRevLett.116.173002}. We assume the probe laser is weak, all the lasers and the MW fields are resonant with the atomic transitions $\Delta_{p}=\Delta_{d}=\Delta_{c}=\Delta_{MW}=0$.
The atoms are initially populated in the ground state. The reduced optical Bloch equation of the atoms is given based on the first-order perturbation of the probe light. The power (strength) of the beat note against its frequency is calculated through the quasi-steady-state solution of equations below:
}
\begin{equation}
\begin{aligned}\frac{d}{dt}\rho_{21} & =-\frac{1}{2}(\Gamma_{e}+2\gamma)\rho_{21}+\frac{i}{2}\left(\Omega_{p}+\Omega_{d}\rho_{31}\right)\\
\frac{d}{dt}\rho_{31} & =-\left(2\Gamma_{d}+2\gamma\right)\rho_{31}+\frac{i}{2}\left(\Omega_{d}\rho_{21}+\Omega_{c}\rho_{41}\right)\\
\frac{d}{dt}\rho_{41} & =-\frac{1}{2}\left(2\gamma+\Gamma_{r1}\right)\rho_{41}+\frac{i}{2}(\Omega_{c}\rho_{31}+\Omega_{MW}(t)\rho_{51})\\
\frac{d}{dt}\rho_{51} & =-\frac{1}{2}\left(2\gamma+\Gamma_{r2}\right)\rho_{51}+\frac{i}{2}\Omega_{MW}(t)\rho_{41}
\end{aligned}
\end{equation}
\label{app:theorycal}
where, the $\gamma$ represent the transit dephasing rate of the ground state. With $\gamma\gg\Gamma_{r1},\Gamma_{r2}$, the decay terms from the Rydberg state the high-order derivative of the $\rho_{21}$ are ignored. This transient pattern (the instant bandwidth) is independent of the LO frequency.  The experimental frequency response data shows a good agreement with our bandwidth model calculation (black solid line) in Fig.~\ref{fig2}(a). And the curve is also fit
well with an exponential decay function. We use a scaling factor to our theory curves which only considers the relative intensity of the beat note, as we set the maximum signal amplitude as a reference point.
\label{app:theorycal}
\section{Dynamic range variation with the MFC lines number}

The sensitivity of the system is dependent on the number of the MFC comb lines and the power of the MFC field. We plot the power of the detected beat note versus the $E$-field intensity of the applied signal  when using a single LO field in Fig.~\ref{fig8}(a) for comparison. 
Fig.~\ref{fig8}(b) shows the beat-note response when using a MFC field with 3 comb lines and a repetition rate of $f_{r}=300\,$ kHz. We want to emphasize that the dynamic range is affected by the relative phase of each line of the MFC. There is a difference for dynamic range of $\sim$8\,dB (from 65\,dB to 57\,dB) between the MFC with optimized phase and the one without optimization. The optimized phase of the each frequency line is set as $0$, $0$, and $\pi$.  We only plot input signals have frequency offsets $\delta f$ of 50\,kHz, 350\,kHz from the centre frequency ($f_{LO}\ensuremath{=4.485}$\,GHz). To compare the response within the covering range of different comb lines, the beat-note frequencies are all 50\,kHz. Fig.~\ref{fig8}(c) (Fig.~\ref{fig8}(d)) shows the response when the MFC with 5 (7) comb lines are set as LO, the average dynamic range is 62.8\,dB (56\,dB). According to the Eq.~\ref{eq}, the beat note response decrease with the increase of the number of the MFC lines.
\bibliography{ref}
\end{document}